\documentclass[10pt]{article}

\usepackage{setspace}
\usepackage{amsmath}%
\usepackage{amsfonts}%
\usepackage{amssymb}%
\usepackage{graphicx}
\usepackage{microtype}

\begin{document}
\onehalfspacing
\raggedbottom

\title{Deriving Proper Uniform Priors for Regression Coefficients, Part II}
\author{H.R.N.~van~Erp, R.O.~Linger, and P.H.A.J.M.~van~Gelder}
\date{}

\maketitle

\begin{abstract}
\noindent It is a relatively well-known fact that in problems of Bayesian model selection improper priors should, in general, be avoided. In this paper we derive a proper and parsimonious uniform prior for regression coefficients. We then use this prior to derive the corresponding evidence values of the regression models under consideration. By way of these evidence values one may proceed to compute the posterior probabilities of the competing regression models.
\end{abstract}

\section{Introduction}
\noindent We, that is, the authors of this article, were in a position that we had to select from a considerable number spline models, that is, highly variate regression models. As these spline models may have hundreds of regression coefficients, we were forced to think about the most suitable bounds of the non-informative priors of the unknown parameters. Not because this would give us better parameter estimates, but simply because taking a uniform prior with overly large bounds would severely punish the larger regression models. 

Grappling with this problem, we ended up with a uniform prior for the regression coefficients $\beta$, which is derived by putting $k$-sigma bounds on $\left\|\mathbf{e}\right\|$, that is, the length of the error vector $\mathbf{e}$. Note that it is the multivariate probability distribution which we assign to the error vector that allows us to construct the likelihood function for some output vector $\mathbf{y}$. But, as it would seem, this multivariate probability distribution may also guide us in the construction of a parsimonious prior distribution for the unknown regression coefficients. 

The evidence value, which results from this parsimonious prior is analytical and has as its sufficient statistics the sample size, $N$, the number of parameters used, $m$, the goodness of fit, $\left\|\mathbf{y} - \hat{\mathbf{y}}\right\|$, and the sigma bound on the length of the error vector, $k$. 

The structure of this paper is as follows. First we give a quick overview of the basic constructs of Bayesian statistics. Then we discuss the role of the evidence construct in Bayesian model selection. We then proceed to give a Bayesian regression analysis for the case where the spread $\sigma$ is assumed to be known. This provides the pertinent context for the probabilistic prior of the unknown regression coefficients. We then proceed to derive the probabilistic prior and the corresponding evidence value, for the case where the spread $\sigma$ is assumed to be known. We then use the probabilistic prior to compute the evidence value for the case, typically encountered in practice, where the spread $\sigma$ is unknown. Finally, for completeness' sake, we give the Bayesian regression analysis for the case where the spread $\sigma$ is assumed to be unknown.

\section{Bayesian statistics}
Bayesian statistics has four fundamental constructs, namely, the prior, the likelihood, the posterior, and the evidence. These constructs are related in the following way:
\begin{equation}
	\label{eq1}
	\text{posterior} = \frac{\text{prior} \times \text{likelihood}}{\text{evidence}}
\end{equation}
Most of us will be intimately familiar with the prior, likelihood, and posterior. However, the evidence concept is less universally known, as most people come to Bayesianity by way of the more compact relationship
\begin{equation}
	\label{eq1b}
	\text{posterior} \propto \text{prior} \times \text{likelihood}
\end{equation}
which does not make any explicit mention of the evidence construct; see for example \cite{Zellner71} throughout. 

In what follows, we will employ in our analyses the correct, though notationally more cumbersome, relation \eqref{eq1}, and forgo of the more compact, but incomplete, Bayesian shorthand \eqref{eq1b}. This is done so the reader may develop some feel for the evidence construct, and how this construct relates to the other three Bayesian constructs of prior, likelihood, and posterior. 

Let $p\left(\left.\theta\right|I\right)$ be the prior of some parameter $\theta$, where $I$ is the prior information regarding the unknown $\theta$ which we have to our disposal. Let $p\left(\left.D\right|\theta, M\right)$ be the probability of the data $D$ conditional on the value of parameter $\theta$ and the likelihood model $M$ which is used; the probability of the data is also known as the likelihood of the parameter $\theta$. Let $p\left(\left.\theta\right|D, M, I\right)$ be the posterior distribution of the parameter $\theta$, conditional on the data $D$, the likelihood model $M$, and the prior model information $I$. Then
\begin{equation}
	\label{eq2}
	p\left(\left.\theta\right|D, M, I\right)  =  \frac{p\left(\left.\theta, D\right|M, I\right)}{ p\left(\left.D\right|M, I\right)} =  \frac{p\left(\left.\theta\right|I\right) p\left(\left.D\right|\theta, M\right)}{\int p\left(\left.\theta\right|I\right) p\left(\left.D\right|\theta, M\right) d\theta}
\end{equation}
where
\begin{equation}
	\label{eq3}
p\left(\left.D\right|M, I\right) = \int p\left(\left.\theta, D\right|M, I\right) d\theta = \int p\left(\left.\theta\right|I\right) p\left(\left.D\right|\theta, M\right) d\theta 
\end{equation}
is the evidence, that is, marginalized likelihood of both the likelihood model $M$ and the prior information model $I$. In the next section we will show how the evidence is used in Bayesian model selection.

\section{Bayesian model selection}
If we have a set of likelihood models $M_{j}$ we wish to choose from, and just the one prior information model $I$, then we may do so by computing the evidence values $p\left(\left.D\right|M_{j}, I\right)$. Let $p\left(M_{j}\right)$ and $p\left(\left.M_{j}\right|D, I\right)$ be, respectively, the prior and posterior probability of the likelihood model $M_{j}$. Then the posterior probability distribution of these likelihood models is given as
\begin{equation}
	\label{eq4}
p\left(\left.M_{j}\right|D, I\right) = \frac{p\left(M_{j}\right) p\left(\left.D\right|M_{j}, I\right)}{\sum_{j} p\left(M_{j}\right) p\left(\left.D\right|M_{j}, I\right)}
\end{equation}
Note that if $p\left(M_{j}\right) = p\left(M_{k}\right)$ for  all $j$ and  $k$, then we have that \eqref{eq4} reduces to
\begin{equation}
	\label{eq5}
p\left(\left.M_{j}\right|D, I\right) = \frac{p\left(\left.D\right|M_{j}, I\right)}{\sum_{j} p\left(\left.D\right|M_{j}, I\right)}
\end{equation}
Stated differently, if we assign equal prior probabilities to our different likelihood models, the posterior probabilities of these models reduce to their normalized evidence values, that is, the models may be ranked by their respective evidence values \cite{MacKay03}.

We also may have the situation in which we have a set of prior information models to choose from. For example, in image reconstruction we have that all the artfulness goes into the construction of an informative prior, whereas the likelihood model is trivial and remains the same for all prior models considered, see for example \cite{Skilling91}. Let $p\left(\left.D\right|M, I_{j}\right)$ be the evidence values of the prior information model $I_{j}$, and let $p\left(I_{j}\right)$ and $p\left(\left.I_{j}\right|D, M\right)$, respectively, be their prior and posterior probabilities. Then the posterior probability distribution of the prior information models is given as
\begin{equation}
	\label{eq6a}
p\left(\left.I_{j}\right|D, M\right) = \frac{p\left(I_{j}\right) p\left(\left.D\right|M, I_{j}\right)}{\sum_{j} p\left(I_{j}\right) p\left(\left.D\right|M, I_{j}\right)}
\end{equation}
And again, if $p\left(I_{j}\right) = p\left(I_{k}\right)$ for all $j$ and $k$, we have that the prior information models may be ranked by their respective evidence values: 
\begin{equation}
	\label{eq6}
p\left(\left.I_{j}\right|D, M\right) = \frac{p\left(\left.D\right|M, I_{j}\right)}{\sum_{j} p\left(\left.D\right|M, I_{j}\right)}
\end{equation}

In model selection for Bayesian regression analyses, we have yet another scenario, in which both the likelihood model, $M_{j}$, and the corresponding prior model, $I_{j}$, are determined by the particular choice of the $N \times m$ predictor matrix $X$ (as will be demonstrated in this paper). Let $p\left(\left.D\right|M_{j} I_{j}\right)$ be the evidence value of the ensemble of the prior information and likelihood model, that is, $I_{j} M_{j}$, and let $p\left(I_{j} M_{j}\right)$ and $p\left(\left.I_{j} M_{j}\right|D\right)$, respectively, be their prior and posterior probabilities. Then the posterior probability distribution of these ensembles is given as
\begin{equation}
	\label{eq6b}
p\left(\left.I_{j} M_{j}\right|D\right) = \frac{p\left(I_{j} M_{j}\right) p\left(\left.D\right|I_{j} M_{j}\right)}{\sum_{j} p\left(I_{j} M_{j}\right) p\left(\left.D\right|I_{j} M_{j}\right)}
\end{equation}
Again, if $p\left(I_{j} M_{j}\right) = p\left(I_{k} M_{k}\right)$ for  all $j$ and  $k$, we have that the ensemble of the prior information and likelihood models may be ranked by their respective evidence values:
\begin{equation}
	\label{eq6c}
p\left(\left.I_{j} M_{j}\right|D\right) = \frac{p\left(\left.D\right|I_{j} M_{j}\right)}{\sum_{j} p\left(\left.D\right|I_{j} M_{j}\right)}
\end{equation}  

Note that the right-hand sides of \eqref{eq5}, \eqref{eq6}, \eqref{eq6c} all pertain to the same scaled evidence values, though their left-hand sides refer to different posteriors. Consequently, scaled evidences may be many different things to many different people, depending on the context of their analyses. 

\section{Bayesian regression analysis for known $\sigma$}
Let the model $M$ for the output vector $\mathbf{y}$ be,
\begin{equation}
	\label{eq7}
	\mathbf{y} = X \beta + \mathbf{e}
\end{equation}
where $X$ is some $N \times m$ predictor matrix, $\beta$ is the $m \times 1$ vector with regression coefficients, and $\mathbf{e}$ is the $N \times 1$ error vector to which we assign the multivariate normal distribution:
\begin{equation}
	\label{eq7b}
	p\left(\left.\mathbf{e}\right| \sigma\right) = \frac{1}{\left(2 \pi \sigma^{2}\right)^{N/2}} \exp\left(-\frac{\mathbf{e}^{T}\mathbf{e}}{2\sigma^{2}} \right)
\end{equation}
or, equivalently, $\mathbf{e} \sim MN\left(\mathbf{0}, \sigma^{2} I\right)$, where $I$ is the $N \times N$ identity matrix and $\sigma$ is some known standard deviation

By way of a simple Jacobian transformation from $\mathbf{e}$ to $\mathbf{y}$\footnote{The transformation $\mathbf{e} = \mathbf{y} - X \beta$ has a corresponding Jacobian of unity, that is, $J = 1$.}, \eqref{eq7} and \eqref{eq7b}, we construct the likelihood function:
\begin{equation}
	\label{eq8}
	p\left(\left.\mathbf{y}\right|\sigma, X, \beta, M\right) = \frac{1}{\left(2 \pi \sigma^{2}\right)^{N/2}} \exp\left[-\frac{1}{2\sigma^{2}} \left(\mathbf{y} - X \beta\right)^{T}\left(\mathbf{y} - X \beta\right)\right]
\end{equation}
We assign a uniform prior to the unknown regression coefficients $\beta$, \cite{Zellner71},
\begin{equation}
	\label{eq9}
	p\left(\left.\beta\right| I\right) = C
\end{equation}
where $C$, is a yet unspecified constant and $I$ is the prior information regarding the unknown $\beta$'s, which we have at our disposal. By way of the Bayesian product rule, see also \eqref{eq2} and \eqref{eq3},
\[
		P\!\left(\left.A B\right|I\right) = P\!\left(\left.A\right|I\right) P\!\left(\left.B\right|A\right) 
\]
we may derive the probability distribution of both vectors $\beta$ and $\mathbf{y}$
\begin{align}
	\label{eq10}
	p\left(\left.\beta,\mathbf{y}\right|\sigma, X, M, I\right) &=  p\left(\left.\beta\right| I\right) p\left(\left.\mathbf{y}\right| X, \beta, \sigma, M\right) \nonumber \\
	 \\
	&= \frac{C}{\left(2 \pi \sigma^{2}\right)^{N/2}} \exp\left[-\frac{1}{2\sigma^{2}} \left(\mathbf{y} - X \beta\right)^{T}\left(\mathbf{y} - X \beta\right)\right] \nonumber 
\end{align}
By integrating the unknown $\beta$ out of \eqref{eq10}, we obtain the evidence of both $M$ and $I$, \eqref{eq3}: 
\begin{equation}
	\label{eq11}
	p\left(\left.\mathbf{y}\right|\sigma, X, M, I\right) = \int p\left(\left.\beta,\mathbf{y}\right| X, \sigma, M, I\right) d\beta
\end{equation}
The evidence \eqref{eq11} is used both to normalize \eqref{eq10} into a posterior distribution, by way of the relation \eqref{eq2},
as well as to choose between competing regression models, \eqref{eq5}.

In order to evaluate the evidence \eqref{eq11}, we may rewrite the inner vector product in the exponential of \eqref{eq10} as, Appendix A,
\begin{equation}
	\label{eq12}
	\left(\mathbf{y} - X \beta\right)^{T}\left(\mathbf{y} - X \beta\right) = \left(\mathbf{y}-\hat{\mathbf{y}}\right)^{T}\left(\mathbf{y}-\hat{\mathbf{y}}\right) + \left(\beta - \hat{\beta}\right)^{T}X^{T}X\left(\beta - \hat{\beta}\right) 
\end{equation}
where
\begin{equation}
	\label{eq13}	
	\hat{\beta} = \left(X^{T}X\right)^{-1} X^{T}\mathbf{y}
\end{equation}
and 
\begin{equation}
	\label{eq14}
		\hat{\mathbf{y}} = X \hat{\beta} = X\left(X^{T}X\right)^{-1} X^{T}\mathbf{y}
\end{equation}
Substituting the decomposition \eqref{eq12} into \eqref{eq10}, we obtain
\[
	p\left(\left.\beta,\mathbf{y}\right|\sigma, X, M, I\right) = \frac{C}{\left(2 \pi \sigma^{2}\right)^{N/2}} \exp\left\{-\frac{1}{2\sigma^{2}} \left[\left(\mathbf{y}-\hat{\mathbf{y}}\right)^{T}\left(\mathbf{y}-\hat{\mathbf{y}}\right) + \left(\beta - \hat{\beta}\right)^{T}X^{T}X\left(\beta - \hat{\beta}\right)\right]\right\} 
\]
which may be factored as
\begin{align}
	\label{eq18}
	p\left(\left.\beta,\mathbf{y}\right|\sigma, X, M, I\right) &= \frac{C}{\left|X^{T}X\right|^{1/2}\left(2 \pi \sigma^{2}\right)^{\left(N-m\right)/2}} \exp\left[-\frac{1}{2\sigma^{2}} \left(\mathbf{y}-\hat{\mathbf{y}}\right)^{T}\left(\mathbf{y}-\hat{\mathbf{y}}\right)\right] \nonumber \\
	\nonumber \\
	&\qquad \times \frac{\left|X^{T}X\right|^{1/2}}{\left(2 \pi \sigma^{2}\right)^{m/2}}\exp\left[-\frac{1}{2\sigma^{2}}\left(\beta - \hat{\beta}\right)^{T}X^{T}X\left(\beta - \hat{\beta}\right)\right] \nonumber \\
\end{align}
The last term in \eqref{eq18} evaluates to 1 when integrated over the $\beta$ vector, as it is in the multivariate normal form, \cite{Zellner71}. Consequently, we have, by way of the factorization \eqref{eq18}, that the evidence, that is, integral \eqref{eq11}, evaluates to
\begin{equation}
	\label{eq19}
	p\left(\left.\mathbf{y}\right|\sigma, X, M, I\right) = \frac{C}{\left|X^{T}X\right|^{1/2}\left(2 \pi \sigma^{2}\right)^{\left(N-m\right)/2}} \exp\left[-\frac{1}{2\sigma^{2}} \left(\mathbf{y}-\hat{\mathbf{y}}\right)^{T}\left(\mathbf{y}-\hat{\mathbf{y}}\right)\right] 
\end{equation}
 If we then substitute \eqref{eq18} and \eqref{eq19} into \eqref{eq2}, we obtain the posterior of the unknown $\beta$ vector, 
\begin{align}
	\label{eq20}
	p\left(\left.\beta\right|\sigma, \mathbf{y}, X, M, I\right) &= \frac{p\left(\left.\beta,\mathbf{y}\right| \sigma, X, M, I\right)}{p\left(\left.\mathbf{y}\right|\sigma, X, M, I\right)}\nonumber \\
	\nonumber \\
	&= \frac{\left|X^{T}X\right|^{1/2}}{\left(2 \pi \sigma^{2}\right)^{m/2}}\exp\left[-\frac{1}{2\sigma^{2}}\left(\beta - \hat{\beta}\right)^{T}X^{T}X\left(\beta - \hat{\beta}\right)\right] \nonumber \\
\end{align}
It can be seen  that posterior of the unknown $\beta$ has a mean of, \eqref{eq13}, $\hat{\beta} = \left(X^{T}X\right)^{-1} X^{T}\mathbf{y}$, and a covariance matrix of $\left(X^{T}X/\sigma^{2}\right)^{-1}$.

Note that in parameter estimation problem, that is, the derivation of the posterior distribution \eqref{eq20}, all reference to the uniform prior $C$, \eqref{eq9}, has fallen away. In contrast, in the model selection problem, that is, the derivation of the evidence \eqref{eq19}, $C$ is still very much there. 

\section{Assigning a parsimonious prior}
\noindent We now try to specify the constant $C$ in the prior \eqref{eq9}. By way of \eqref{eq7}, we have that for a $N \times m$ predictor matrix $X$ or rank $m$, 
\begin{equation}
	\label{eq21}
	\beta =  \left(X^{T}X\right)^{-1} X^{T}\left(\mathbf{y} - \mathbf{e}\right) = \hat{\beta} - \left(X^{T}X\right)^{-1} X^{T}\mathbf{e} 
\end{equation}
where $\mathbf{e} \sim MN\left(\mathbf{0}, \sigma^{2} I\right)$, \eqref{eq7b}. Closer inspection of \eqref{eq21} shows us that the parameter space of $\beta$ is a-priori constrained by the error vector $\mathbf{e}$. We will now demonstrate this for the special case where the predictor matrix $X$ is a $N \times 1$ vector $\mathbf{x}$. 

By way of \eqref{eq21}, we have that
\begin{equation}
	\label{eq22a}
 \beta = \hat{\beta} - \frac{\mathbf{x}^{T}\mathbf{e}}{\mathbf{x}^{T}\mathbf{x}} = \hat{\beta} -  \cos\theta \frac{\left\|\mathbf{x}\right\| \left\|\mathbf{e}\right\|}{\left\|\mathbf{x}\right\|^{2} } 
\end{equation}
where $\theta$ is the angle between the predictor vector $\mathbf{x}$ and the error vector $\mathbf{e}$, $\left\|\mathbf{x}\right\|$ is the length of $\mathbf{x}$, $\left\|\mathbf{e}\right\|$ is the length of $\mathbf{e}$. Seeing that $-1 \leq \cos\theta \leq 1$, we may by way of \eqref{eq22a} put definite bounds on $\beta$
\begin{equation}
	\label{eq22}
 \hat{\beta} -  \frac{\max \left\|\mathbf{e}\right\|}{\left\|\mathbf{x}\right\|}  \leq \beta \leq \hat{\beta} +  \frac{\max \left\|\mathbf{e}\right\|}{\left\|\mathbf{x}\right\|} 
\end{equation}
Stated differently, if we assign a uniform distribution to the regression coefficient $\beta$, then this uniform distribution is defined on a line-piece of length $2 \max \left\|\mathbf{e}\right\|/\left\|\mathbf{x}\right\|$, and it follows that, for the case of just the one regression coefficient, the prior \eqref{eq9} can be derived to be   
\begin{equation}
	\label{eq23}
	p\left(\left.\beta\right| I\right) = \frac{\left\|\mathbf{x}\right\|}{2\max \left\|\mathbf{e}\right\|}
\end{equation}
where \eqref{eq23}, is understood to be centered on $\hat{\beta}$. 

In Appendix B it is demonstrated that for the case where $X$ is a $N \times m$ predictor matrix, \eqref{eq22} and \eqref{eq23} generalize to the statements that $\beta$ is constrained to lie in an $m$-dimensional ellipsoid which is centered on $\hat{\beta}$ and has a volume of
\begin{equation}
	\label{eq24}
	V = \frac{\pi^{m/2}}{\Gamma\left[\left(m+2\right)/2\right]} \frac{\left(\max \left\|\mathbf{e}\right\|\right)^{m}}{\left|X^{T}X\right|^{1/2}}
\end{equation}
and that the corresponding multivariate uniform prior is the inverse of this volume:
\begin{equation}
	\label{eq25}
	p\left(\left.\beta\right| I\right) = \frac{\Gamma\left[\left(m+2\right)/2\right]}{\pi^{m/2}} \frac{\left|X^{T}X\right|^{1/2}}{\left(\max \left\|\mathbf{e}\right\|\right)^{m}}
\end{equation}
where \eqref{eq25} is understood to be centered on $\hat{\beta}$. 

Seeing that $\mathbf{e}$ has a known probability distribution, \eqref{eq7b}, 
\[
	p\left(\left.\mathbf{e}\right| \sigma\right) = \frac{1}{\left(2 \pi \sigma^{2}\right)^{N/2}} \exp\left(-\frac{\mathbf{e}^{T}\mathbf{e}}{2\sigma^{2}} \right)
\]
we may derive, by way of a Jacobian transformation, the marginal probability distribution of $\left\|\mathbf{e}\right\|$, that is, the length of the error vector $\mathbf{e}$, Appendix B: 
\begin{equation}
	\label{eq27}
	p\left(\left.\left\|\mathbf{e}\right\|\right| \sigma\right) = \frac{2 \: \left\|\mathbf{e}\right\|^{N-1}}{\left(2 \sigma^{2}\right)^{N/2} \Gamma\left(N/2\right)} \exp\left(-\frac{\left\|\mathbf{e}\right\|^2}{2\sigma^{2}} \right)
\end{equation}
This probability distribution has a mean 
\begin{equation}
	\label{eq28}
	E\left(\left\|\mathbf{e}\right\|\right) = \frac{\sqrt{2} \:\Gamma\left[\left(N + 1\right)/2\right]}{\Gamma\left(N/2\right)} \sigma \approx \sqrt{N - 1} \; \sigma
\end{equation}
and a variance
\begin{equation}
	\label{eq29}
	\text{var}\left(\left\|\mathbf{e}\right\|\right) = \left(N - \left\{\frac{\sqrt{2} \: \Gamma\left[\left(N + 1\right)/2\right]}{\Gamma\left(N/2\right)}\right\}^{2}\right) \sigma^{2} \approx \sigma^{2}
\end{equation}
By way \eqref{eq28} and \eqref{eq29}, we may give a probabilistic interpretation of $\max \left\|\mathbf{e}\right\|$ in \eqref{eq25}, that is, we let
\begin{equation}
	\label{eq30}
	\max \left\|\mathbf{e}\right\| = E\left(\left\|\mathbf{e}\right\|\right) + k \sqrt{\text{var}\left(\left\|\mathbf{e}\right\|\right)} \approx \left(\sqrt{N - 1} + k\right)  \; \sigma
\end{equation}
where $k$ is some suitable sigma upper bound, for example, $k = 6$. Note, that for small sample sizes $N$ one should be careful to use in \eqref{eq30} the exact terms of \eqref{eq28} and \eqref{eq29}, as opposed to their approximations. 

In what follows, we will assume large sample sizes $N$ and, consequently, stick with the simpler approximation \eqref{eq30}. Substituting \eqref{eq30} into \eqref{eq25}, we obtain the prior of the $\beta$'s we are looking for
\begin{equation}
	\label{eq31}
	p\left(\left.\beta\right|\sigma, I\right) \approx \frac{\Gamma\left[\left(m+2\right)/2\right]}{\pi^{m/2}} \frac{\left|X^{T}X\right|^{1/2}}{\left[\left(\sqrt{N - 1} + k\right)\sigma\right]^{m}}
\end{equation}
Note that the prior \eqref{eq31} is conditional upon the spread parameter $\sigma$. 

By way of \eqref{eq9}, we may substitute \eqref{eq31} into \eqref{eq19}, and so obtain the evidence value of the likelihood model $M$ and prior information $I$, conditional on some known $\sigma$,
\begin{equation}
	\label{eq32}
	p\left(\left.\mathbf{y}\right|\sigma, X, M, I\right) \approx \frac{2^{m/2} \: \Gamma\left[\left(m+2\right)/2\right]}{\left(\sqrt{N - 1} + k\right)^{m}} \frac{1}{\left(2 \pi \sigma^{2}\right)^{N/2}} \exp\left[-\frac{1}{2\sigma^{2}} \left(\mathbf{y}-\hat{\mathbf{y}}\right)^{T}\left(\mathbf{y}-\hat{\mathbf{y}}\right)\right] 
\end{equation}

\section{The evidence for unknown $\sigma$}
By assigning the Jeffreys prior 
\begin{equation}
	\label{eq33}
	p\left(\sigma\right) = \frac{A}{\sigma}
\end{equation}
where $A$ is some normalizing constant, to the evidence \eqref{eq32}, we may integrate out the unknown $\sigma$, see also \eqref{eq3},
\begin{equation}
	\label{eq34}
	p\left(\left.\mathbf{y}\right|X, M, I\right) = \int p\left(\left.\sigma,\mathbf{y}\right|X, M, I\right) d\sigma = \int p\left(\sigma\right) p\left(\left.\mathbf{y}\right|\sigma, X, M, I\right) d\sigma
\end{equation}
where, \eqref{eq32}, \eqref{eq33}, and \eqref{eq34},
\begin{equation}
	\label{eq35}
	p\left(\left.\sigma,\mathbf{y}\right|X, M, I\right) \approx \frac{2^{m/2} \: \Gamma\!\left[\left(m+2\right)/2\right]}{\left(\sqrt{N - 1} + k\right)^{m}} \frac{A}{\left(2 \pi\right)^{N/2} \sigma^{N + 1}} \exp\left[-\frac{1}{2\sigma^{2}} \left(\mathbf{y}-\hat{\mathbf{y}}\right)^{T}\left(\mathbf{y}-\hat{\mathbf{y}}\right)\right] 
\end{equation}
We may conveniently factorize \eqref{eq35} as, 
\begin{align}
	\label{eq36}
	p&\left(\left.\sigma,\mathbf{y}\right|X, M, I\right) \approx \frac{2^{m/2} \: \Gamma\!\left[\left(m+2\right)/2\right]}{\left(\sqrt{N - 1} + k\right)^{m}} \frac{1}{\left\|\mathbf{y}-\hat{\mathbf{y}}\right\|^{N}} \frac{A\:\Gamma\!\left(N/2\right)}{2 \pi^{N/2}} \nonumber \\
\nonumber \\
	&\times \frac{2}{\Gamma\!\left(N/2\right)}\left(\frac{\left\|\mathbf{y}-\hat{\mathbf{y}}\right\|^{2}}{2}\right)^{N/2} \frac{1}{\sigma^{N + 1}} \exp\left[-\frac{1}{2\sigma^{2}} \left(\mathbf{y}-\hat{\mathbf{y}}\right)^{T}\left(\mathbf{y}-\hat{\mathbf{y}}\right)\right] \nonumber  \\
\end{align}
where last term in \eqref{eq36} evaluates to 1 when integrated over $\sigma$; as it has the form of an inverted gamma distribution, \cite{Zellner71}. Consequently, we have, by way of the factorization \eqref{eq36}, that the evidence, that is, the integral \eqref{eq34}, evaluates to
\begin{equation}
	\label{eq37}
	p\left(\left.\mathbf{y}\right|X, M, I\right) \approx \frac{2^{m/2} \: \Gamma\!\left[\left(m+2\right)/2\right]}{\left(\sqrt{N - 1} + k\right)^{m}} \frac{1}{\left\|\mathbf{y}-\hat{\mathbf{y}}\right\|^{N}} \frac{A\:\Gamma\!\left(N/2\right)}{2 \pi^{N/2}}
\end{equation}
The evidence \eqref{eq37} consists of an Occam Factor, which penalizes the number of parameters and which is a monotonic decreasing function in $m$:
\begin{equation}
	\label{eq38}
	\text{\textit{Occam Factor}} = \frac{2^{m/2} \: \Gamma\!\left[\left(m+2\right)/2\right]}{\left(\sqrt{N - 1} + k\right)^{m}} 
\end{equation}
a goodness-of-fit factor, which rewards a good fit of the likelihood model $M$:
\begin{equation}
	\label{eq39}
	\text{\textit{Goodness-of-Fit}} = \frac{1}{\left\|\mathbf{y}-\hat{\mathbf{y}}\right\|^{N}} 
\end{equation}
and a common factor
\[
	\text{\textit{Common Factor}} = \frac{A\:\Gamma\!\left(N/2\right)}{2 \pi^{N/2}}
\]
which is a shared by all evidence values and which cancels out as the posterior probabilities of the models are computed, \eqref{eq4}. 

Note that the analytical evidence \eqref{eq37} has as its sufficient statistics the sample size, $N$, the number of parameters used, $m$, the goodness of fit $\left\|\mathbf{y}-\hat{\mathbf{y}}\right\|$, and the sigma bound $k$. 

\section{The Posterior of $\beta$ for Unknown $\sigma$}
We now will, for completeness sake derive the posterior of the $\beta$'s, which is associated with the evidence value \eqref{eq37}. We assign as priors for the $\beta$'s and $\sigma$, \eqref{eq31} and \eqref{eq33},
\begin{equation}
	\label{eq40}
		p\left(\left.\sigma,\beta\right|I\right) = p\left(\sigma\right)   p\left(\left.\beta\right|\sigma,I\right) \approx \frac{A}{\sigma} \frac{\Gamma\left[\left(m+2\right)/2\right]}{\pi^{m/2}} \frac{\left|X^{T}X\right|^{1/2}}{\left(\sqrt{N - 1} + k\right)^{m} \sigma^{m}} 
\end{equation}
Multiplying the prior \eqref{eq37} with the likelihood \eqref{eq8}, and dividing by the evidence \eqref{eq37}, we obtain the posterior of $\beta$ and $\sigma$,
\begin{equation}
	\label{eq41}
	p\left(\left.\sigma,\beta\right|X, \mathbf{y}, M, I\right) \approx \frac{\left\|\mathbf{y}-\hat{\mathbf{y}}\right\|^{N}}{\sigma} \frac{2 \pi^{N/2}}{\Gamma\!\left(N/2\right)}  \frac{\left|X^{T}X\right|^{1/2}}{\left(2 \pi \sigma^{2}\right)^{\left(N+m\right)/2}} \exp\left[-\frac{1}{2\sigma^{2}} \left(\mathbf{y} - X \beta\right)^{T}\left(\mathbf{y} - X \beta\right)\right]
\end{equation}
The marginalized posterior of the $\beta$ vector is 
\begin{equation}
	\label{eq42}
	p\left(\left.\beta\right|X, \mathbf{y}, M, I\right)  = \int p\left(\left.\sigma,\beta\right|X, \mathbf{y}, M, I\right) d\sigma
\end{equation}
We may factor \eqref{eq41} as
\begin{align}
	\label{eq43}
	p&\left(\left.\sigma,\beta \right| X, \mathbf{y}, M, I\right) \approx \frac{\Gamma\!\left[\left(N+m\right)/2\right]}{\Gamma\!\left(N/2\right)} \frac{\left|X^{T}X\right|^{1/2}}{\pi^{m/2}} \frac{\left\|\mathbf{y}-\hat{\mathbf{y}}\right\|^{N}}{\left\|\mathbf{y} - X \beta\right\|^{N+m}} \nonumber\\
	\nonumber\\
	&\times \frac{2}{\Gamma\!\left[\left(N+m\right)/2\right]}\left(\frac{\left\|\mathbf{y} - X \beta\right\|^{2}}{2}\right)^{\left(N+m\right)/2}\frac{1}{\sigma^{N+m+1}}\exp\left(-\frac{1}{2\sigma^{2}} \left\|\mathbf{y} - X \beta\right\|^{2}\right)
\end{align}
where the last term in \eqref{eq43} evaluates to 1 when integrated over $\sigma$; as it has the form of an inverted gamma distribution, \cite{Zellner71}. Consequently, we have, by way of the factorization \eqref{eq43}, that the marginalized posterior of $\beta$, that is, the integral \eqref{eq42} evaluates to a multivariate Student-t distribution, \cite{Zellner71}:
\begin{equation}
	\label{eq44}
	p\left(\left.\beta\right|X, \mathbf{y}, M, I\right) \approx \frac{\Gamma\!\left[\left(N+m\right)/2\right] \left|X^{T}X\right|^{1/2} \left\|\mathbf{y}-\hat{\mathbf{y}}\right\|^{N}}{\Gamma\!\left(N/2\right) \pi^{m/2} \left[\left\|\mathbf{y}-\hat{\mathbf{y}}\right\|^{2} + \left(\beta - \hat{\beta}\right)^{T}X^{T}X\left(\beta - \hat{\beta}\right)\right]^{\left(N+m\right)/2}} 
\end{equation}
where we have used \eqref{eq12} to write
\[
		\left\|\mathbf{y} - X \beta\right\|^{2} = \left\|\mathbf{y}-\hat{\mathbf{y}}\right\|^{2} + \left(\beta - \hat{\beta}\right)^{T}X^{T}X\left(\beta - \hat{\beta}\right)
\]
where \eqref{eq13} and \eqref{eq14}, $\hat{\beta} = \left(X^{T}X\right)^{-1} X^{T}\mathbf{y}$ and $\hat{\mathbf{y}} = X \hat{\beta}$. The evidence corresponding with the marginalized posterior \eqref{eq44} is given by \eqref{eq37}.

\section{What is the Data?}
The obvious elephant in the room is the question whether the predictor matrix $X$, used to derive the parsimonious prior \eqref{eq31}, is or is not a part of the data. In \cite{vanErp10} the matrix $X$ was deemed to be part of the data and, consequently, in order to construct the parsimonious prior, one needed to assign a minimum value to the determinant $\left|X^{T}X\right|$, based on the prior information at hand; a non-trivial task. 

This article is a second iteration of the \cite{vanErp10} article, in which it is now suggested that the predictor matrix $X$ is \textit{not} a part of the data. And we offer up two arguments to substantiate this claim. The first argument is that in Bayesian regression analysis the predictor variables $\mathbf{x}_{j}$ are assumed to be `fixed nonstochastic variables', or, alternatively, `random variables distributed independently of the $\mathbf{e}$, with a pdf \textit{not} involving the parameters $\beta_{j}$ and $\sigma$', as stated in \cite{Zellner71}. The second argument, in the same vein, is that the likelihood, that is, the probability distribution of the data, \eqref{eq8}, 
\[
	p\left(\left.\mathbf{y}\right|\sigma, X, \beta, M\right) = \frac{1}{\left(2 \pi \sigma^{2}\right)^{N/2}} \exp\left[-\frac{1}{2\sigma^{2}} \left(\mathbf{y} - X \beta\right)^{T}\left(\mathbf{y} - X \beta\right)\right]
\]
is a probability of $\mathbf{y}$, and not of $X$. This then also would imply that the predictor matrix $X$ should not be considered a part of the data. Rather, $X$ is part of the `prior' problem structure, \eqref{eq7},
\[
	\mathbf{y} = X \beta + \mathbf{e}
\] 
as is the assumed probability distribution of the error vector $\mathbf{e}$.

The benefit of letting $X$ not be a part of the data is that this allows us to derive the parsimonious prior \eqref{eq31}, without having to dub it a `data' prior; a Bayesian oxymoron, if there ever was one.

\section{Discussion}
Using informational consistency requirements, Jaynes \cite{Jaynes68} derived the form of maximal non-informative priors for location parameters, that is, regression coefficients, to be uniform. However, this does not tell us what the limits of this this uniform distribution should be, that is, what particular uniform distribution to use. If we are faced with a parameter estimation problem these limits of the uniform prior are irrelevant, since we may scale the product of the improper uniform prior and the likelihood to one, thus obtaining a properly normalized posterior. However, if we are faced with a problem of model selection then the value of the uniform prior is an integral part of the evidence, which is used to rank the various competing models.We have given here some guidelines for choosing a parsimonious proper uniform prior. To construct such a parsimonious prior one only needs to assign a prior maximal length to the error vector $\mathbf{e}$. In this paper we have treated the case that $\mathbf{e} \sim MN\left(\mathbf{0}, \sigma^{2} I\right)$, both for known and unknown $\sigma$.

\appendix 
\numberwithin{equation}{section}

\section{Decomposing a vector product}
In this appendix we will decompose the inner vector product $\left(\mathbf{y} - X \beta\right)^{T}\left(\mathbf{y} - X \beta\right)$ in the sum of two inner vector products. 

Let
\begin{align}
	\label{eqC12}
\left(\mathbf{y} - X \beta\right)^{T}\left(\mathbf{y} - X \beta\right) &= \mathbf{y}^{T}\mathbf{y} - 2 \mathbf{y}^{T} X \beta + \beta^{T} X^{T}X \beta  \nonumber \\
\nonumber \\
&= \mathbf{y}^{T}\mathbf{y} - 2 \mathbf{y}^{T} X \beta + \beta^{T} X^{T}X \beta + \mathbf{y}^{T}X\left(X^{T}X\right)^{-1} X^{T}\mathbf{y} \nonumber \\
 \\
&\qquad  - \mathbf{y}^{T}X\left(X^{T}X\right)^{-1} X^{T}\mathbf{y} \nonumber
\end{align}
Making use of the identities
\begin{equation}
	\label{eqC13}	
	\hat{\beta} = \left(X^{T}X\right)^{-1} X^{T}\mathbf{y}
\end{equation}
and 
\begin{equation}
	\label{eqC14}
		\hat{\mathbf{y}} = X \hat{\beta} = X\left(X^{T}X\right)^{-1} X^{T}\mathbf{y}
\end{equation}
we have both
\begin{align}
	\label{eqC15}
\left(\mathbf{y}-\hat{\mathbf{y}}\right)^{T}\left(\mathbf{y}-\hat{\mathbf{y}}\right) &= \mathbf{y}^{T}\mathbf{y} - 2 \mathbf{y}^{T}X\left(X^{T}X\right)^{-1} X^{T}\mathbf{y} \nonumber \\
\nonumber \\
& \quad + \mathbf{y}^{T}X\left(X^{T}X\right)^{-1} X^{T}X \left(X^{T}X\right)^{-1} X^{T}\mathbf{y} \nonumber \\
 \\
&= \mathbf{y}^{T}\mathbf{y} - \mathbf{y}^{T}X\left(X^{T}X\right)^{-1} X^{T}\mathbf{y} \nonumber
\end{align}
and 
\begin{align}
	\label{eqC16}
\left(\beta - \hat{\beta}\right)^{T}X^{T}X\left(\beta - \hat{\beta}\right) &= \beta^{T}X^{T}X \beta - 2 \beta^{T}X^{T}X\hat{\beta} + \hat{\beta}^{T} X^{T}X \hat{\beta} \nonumber \\
 \nonumber \\
 &= \beta^{T}X^{T}X \beta - 2 \beta^{T}X^{T}X \left(X^{T}X\right)^{-1} X^{T}\mathbf{y} \nonumber \\
 \\
 &\quad + \mathbf{y}^{T}X\left(X^{T}X\right)^{-1} X^{T}X \left(X^{T}X\right)^{-1} X^{T}\mathbf{y} \nonumber \\
 \nonumber \\
&= \beta^{T}X^{T}X \beta - 2 \mathbf{y}^{T}X^{T}\beta + \mathbf{y}^{T}X\left(X^{T}X\right)^{-1} X^{T}\mathbf{y} \nonumber 
\end{align}
So, by way of \eqref{eqC15} and \eqref{eqC16}, we may rewrite the last right-hand side of \eqref{eqC12} as the sum of two inner vector products
\[
	\left(\mathbf{y} - X \beta\right)^{T}\left(\mathbf{y} - X \beta\right) = \left(\mathbf{y}-\hat{\mathbf{y}}\right)^{T}\left(\mathbf{y}-\hat{\mathbf{y}}\right) + \left(\beta - \hat{\beta}\right)^{T}X^{T}X\left(\beta - \hat{\beta}\right)
\]
This concludes this appendix.

\section{An ellipsoid parameter space}
In this appendix we show that the transformation
\[
		\left(X^{T}X\right)^{-1} X^{T}
\]
will map the vector $\mathbf{e}$ somewhere in an ellipsoid which has a maximal volume of 
\[
	V = \frac{\pi^{m/2}}{\Gamma\left[\left(m+2\right)/2\right]} \frac{\left(\max \left\|\mathbf{e}\right\|\right)^{m}}{\left|X^{T}X\right|^{1/2}}
\]
This result was first derived in \cite{vanErp10}.

Say we have $m$ independent $N \times 1$ vectors $\mathbf{x}_j$ that span some $m$-dimensional orthogonal subspace in the $N$-dimensional data space. We may decompose the vector $\mathbf{e}$ as
\begin{equation}
	\label{eqB1}
	\mathbf{e} = \hat{\mathbf{e}} + \mathbf{n}
\end{equation}
where $\hat{\mathbf{e}}$ is the projection of $\mathbf{e}$ on the $m$-dimensional subspace spanned by the vectors $\mathbf{x}_j$ and $\mathbf{n}$ is the part of $\mathbf{e}$ that is orthogonal to this subspace.

Now, the projection $\hat{\mathbf{e}}$ is mapped on the orthogonal base spanned by the vectors $\mathbf{x}_j$ through the regression coefficients $\beta_{j}$, that is,
\begin{equation}
	\label{eqB2}
	\hat{\mathbf{e}} = \sum_{j=1}^{m} \mathbf{x}_{j} \beta_{j}
\end{equation}
where
\begin{equation}
	\label{eqB3}
	\beta_{j} = \frac{\left\langle\mathbf{x}_{j},\mathbf{e}\right\rangle}{\left\langle\mathbf{x}_{j},\mathbf{x}_{j}\right\rangle} = \frac{\left\langle\mathbf{x}_{j},\hat{\mathbf{e}} + \mathbf{n}\right\rangle}{\left\langle\mathbf{x}_{j},\mathbf{x}_{j}\right\rangle} = \frac{\left\langle\mathbf{x}_{j},\hat{\mathbf{e}} \right\rangle}{\left\langle\mathbf{x}_{j},\mathbf{x}_{j}\right\rangle} = \frac{\left\|\hat{\mathbf{e}}\right\|}{\left\|\mathbf{x}_{j}\right\|} \cos\theta_{j}
\end{equation}	
Because of the independence of the $\mathbf{x}_j$, we have that $\left\langle\mathbf{x}_{i},\mathbf{x}_{j}\right\rangle = 0$, for $i \neq j$. So, if we take the squared norm of \eqref{eqB2} we find, by way of \eqref{eqB3},
\begin{equation}
	\label{eqB4}
	\left\|\hat{\mathbf{e}}\right\|^{2} = \left\|\sum_{j=1}^{m} \mathbf{x}_{j} \beta_{j}\right\|^{2} = \left\|\hat{\mathbf{e}}\right\|^{2} \sum_{j=1}^{m} \cos^{2}\theta_{j}
\end{equation}	
From identity \eqref{eqB4}, it then follows that the angles $\cos\theta_{j}$ in \eqref{eqB3} must obey the constraint
\begin{equation}
	\label{eqB5}
	\sum_{j=1}^{m} \cos^{2}\theta_{j} = 1
\end{equation}	
Combining \eqref{eqB3} and \eqref{eqB5}, we see that all possible values of the coordinates $\beta_{j}$ must lie on the surface of an $m$-variate ellipsoid centered at the origin, having a volume of
\begin{equation}
	\label{eqB6b}
	V = \frac{\pi^{m/2}}{\Gamma\!\left[\left(m+2\right)/2\right]} \prod_{j=1}^{m} r_{j}
\end{equation} 
and with respective axes
\begin{equation}
	\label{eqB6}
	r_{j} = \frac{\left\|\hat{\mathbf{e}}\right\|}{\left\|\mathbf{x}_{j}\right\|}
\end{equation}
Since
\[
	\left\|\hat{\mathbf{e}}\right\| \leq \left\|\mathbf{e}\right\| \leq \max \left\|\mathbf{e}\right\|
\]	
the axes \eqref{eqB6} admit the upper bounds
\begin{equation}
	\label{eqB7}
	\max r_{j} = \frac{\max \left\|\mathbf{e}\right\|}{\left\|\mathbf{x}_{j}\right\|}
\end{equation}	
Consequently, the volume of the parameter space of the $\beta_{j}$ is, for given $\mathbf{x}_{j}$, \eqref{eqB6b},
\begin{equation}
	\label{eqB8}
	V = \frac{\pi^{m/2}}{\Gamma\!\left[\left(m+2\right)/2\right]} \prod_{j=1}^{m} \frac{\max \left\|\mathbf{e}\right\|}{\left\|\mathbf{x}_{j}\right\|}
\end{equation} 
Because of the independence of the $\mathbf{x}_{j}$, we have that the product of the norms $\mathbf{x}_{j}$ is equivalent to the square root of determinant of $X^{T}X$, that is,
\begin{equation}
	\label{eqB9}
	\prod_{j=1}^{m} \left\|\mathbf{x}_{j}\right\| = \left|X^{T}X\right|^{1/2}
\end{equation} 
where $\left|X^{T}X\right|^{1/2}$ is the volume of the parallelepiped defined by the $\mathbf{x}_{j}$. So, we may rewrite as \eqref{eqB8}
\begin{equation}
	\label{eqB10}
	V = \frac{\pi^{m/2}}{\Gamma\!\left[\left(m+2\right)/2\right]} \frac{\left(\max \left\|\mathbf{e}\right\|\right)^{m}}{\left|X^{T}X\right|^{1/2}}
\end{equation} 

If the predictors $\mathbf{x}_{j}$ are not independent, then we may transform them to an orthogonal basis, say, $\tilde{\mathbf{x}}_{j}$, by way of an Gram-Schmidt orthogonalization process. But seeing that the volume of the parallelepiped is invariant under orthogonalization, we have that
\begin{equation}
	\label{eqB11}
	\left|X^{T}X\right|^{1/2} = \left|\tilde{X}^{T}\tilde{X}\right|^{1/2} = \prod_{j=1}^{m} \left\|\tilde{\mathbf{x}}_{j}\right\|
\end{equation} 
where $\tilde{X}$ is the orthogonalized predictor matrix. So, we conclude that \eqref{eqB10} is the volume of the parameter space of the $\beta_{j}$ for both dependent and independent predictors $\mathbf{x}_{j}$.

\section{The probability distribution of $\left\|\mathbf{e}\right\|$}
Let $\mathbf{e} = \left(e_{1},\ldots,e_{N}\right)$ be a $N \times 1$ error vector having the multivariate normal distribution
\begin{equation}
	\label{eqA1}
	p\left(\left.\mathbf{e}\right| \sigma\right) = \frac{1}{\left(2 \pi \sigma^{2}\right)^{N/2}} \exp\left(-\frac{\mathbf{e}^{T}\mathbf{e}}{2\sigma^{2}} \right)
\end{equation}
where $\sigma$ is some known standard deviation. Then we make a change of variable \cite{Zellner71}
\begin{align}
	\label{eqA2}
	\nonumber\\
	e_{1} & = \left\|\mathbf{e}\right\| \cos\alpha_{1} \cos\alpha_{2} \ \cdots \ \cdots \ \cdots \ \cdots \ \cdots \ \cos\alpha_{N-2} \cos\alpha_{N-1} \nonumber \\
	 \nonumber \\
	e_{2} & = \left\|\mathbf{e}\right\| \cos\alpha_{1} \cos\alpha_{2} \ \cdots \ \cdots \ \cdots \ \cdots \ \cdots \ \cos\alpha_{N-2} \sin\alpha_{N-1}\nonumber \\
	&\vdots \nonumber \\
	e_{s} & = \left\|\mathbf{e}\right\| \cos\alpha_{1} \cos\alpha_{2} \ \cdots \ \cos\alpha_{N-s} \sin\alpha_{N-s + 1} \\
	&\vdots \nonumber \\
	e_{N-1} & = \left\|\mathbf{e}\right\| \cos\alpha_{1} \sin\alpha_{2}\nonumber \\
	\nonumber \\
	e_{N} & = \left\|\mathbf{e}\right\| \sin\alpha_{1} \nonumber \\
	\nonumber
\end{align}
The Jacobian of the transformation\eqref{eqA2} is 
\begin{equation}
	\label{eqA3}
	J = \left\|\mathbf{e}\right\|^{N - 1} \cos^{N-2}\alpha_{1} \cos^{N-3}\alpha_{2} \ \cdots \ \cos\alpha_{N-2}
\end{equation}
From trigonometry \eqref{eqA2} yields, 
\begin{equation}
	\label{eqA4}
	\mathbf{e}^{T}\mathbf{e} = \sum_{i = 1}^{N} e_{i}^{2} = \left\|\mathbf{e}\right\|^{2}
\end{equation}
So, substituting \eqref{eqA2}, \eqref{eqA3}, and \eqref{eqA4} into \eqref{eqA1}, we may rewrite the distribution \eqref{eqA1} as
\begin{equation}
	\label{eqA5}
	p\left(\left.\left\|\mathbf{e}\right\|, \alpha_{1}, \ldots, \alpha_{N-1}\right| \sigma\right) = \frac{\left\|\mathbf{e}\right\|^{N-1}}{\left(2 \pi \sigma^{2}\right)^{N/2}} \exp\left(-\frac{\left\|\mathbf{e}\right\|}{2\sigma^{2}} \right) \cos^{N-2}\alpha_{1} \cos^{N-3}\alpha_{2} \ \cdots \ \cos\alpha_{N-2}
\end{equation}
Using, for $j = 1,\ldots, N-2$,
\[
	 \int_{-\pi/2}^{\pi/2} \cos^{N-j-1}\alpha_{j} \; d\alpha_{j} =\pi^{1/2} \frac{\Gamma\!\left[\left(N-j\right)/2\right]}{\Gamma\!\left[\left(N-j-1\right)/2+1\right]}
\]
and, for $j=N-1$,
\[
	 \int_{0}^{2\pi} d\alpha_{N-1} =2 \pi
\]
We are left with the marginal distribution
\begin{equation}
	\label{eqA6}
	p\!\left(\left. \left\|\mathbf{e}\right\| \; \right| \sigma\right) = \frac{2 \left\|\mathbf{e}\right\|^{N-1}}{\left(2 \sigma^{2}\right)^{N/2} \Gamma\!\left(N/2\right)} \exp\left(-\frac{\left\|\mathbf{e}\right\|}{2\sigma^{2}} \right) 
\end{equation}

The $r$th moment of \eqref{eqA6} may be computed by way of the identity
\begin{align}
	\label{eqA7}
	E\!\left(\left\|\mathbf{e}\right\|^{r}\right) &= \int_{0}^{\infty} \left\|\mathbf{e}\right\|^{r} \; p\!\left(\left. \left\|\mathbf{e}\right\| \; \right| \sigma\right)  d\!\left\|\mathbf{e}\right\|  \nonumber \\
	\nonumber \\
	&= \frac{\left(2 \sigma^{2}\right)^{r/2}}{ \Gamma\!\left(N/2\right)}  \int_{0}^{\infty}  \frac{2 \left\|\mathbf{e}\right\|^{N+r-1}}{\left(2 \sigma^{2}\right)^{\left(N+r\right)/2}} \exp\left(-\frac{\left\|\mathbf{e}\right\|}{2\sigma^{2}} \right) d\!\left\|\mathbf{e}\right\|  \nonumber \\
	\nonumber \\
&= \frac{ 2^{r/2} \: \Gamma\!\left[\left(N+r\right)/2\right]}{ \Gamma\!\left(N/2\right)} \sigma^{r} \nonumber
\end{align}

\end{document}